\documentclass[a4paper, 11pt]{article}
\usepackage[titletoc,toc,title]{appendix}
\usepackage{fullpage} % changes the margin

\usepackage{amsmath}
\usepackage{dsfont}
\usepackage{upgreek}
\usepackage{mathtools}
\usepackage{enumitem}

\usepackage{graphicx}
\graphicspath{{./}{./Figures/}}

\usepackage[justification=justified,
            format=plain]{caption}
\usepackage{natbib}
\allowdisplaybreaks

\usepackage{color}
\usepackage{xcolor}

\usepackage{tabularx}
\usepackage{longtable}

\usepackage{floatrow}
\newcommand{\der}[2]{\frac{\mathrm{d}#1}{\mathrm{d}#2}}
\newcommand{\rpar}[1]{\left(#1\right)}
\newcommand{\Hill}[3]{\mathcal{H}_{#1}\left(#2,#3\right)}

\newcommand*{\tabref}[1]{\tablename~\ref{#1}}
\newcommand*{\figref}[1]{\figurename~\ref{#1}}

\renewcommand*{\eqref}[1]{equation~\ref{#1}}

\title{Emergence of regular and complex calcium oscillations by inositol 1,4,5-trisphosphate signaling in astrocytes}
\author{
        Valeri Matrosov, Susan Gordleeva, Natalia Boldyreva \\
            N.I. Lobachevsky State University of Nizhni Novgorod, 603950 Nizhny Novgorod, Russia
        \and
        Eshel Ben-Jacob \\
            School of Physics and Astronomy, Tel Aviv University, 69978 Ramat Aviv, Israel
        \and
        Alexey Semyanov, Victor Kazantsev \\
            N.I. Lobachevsky State University of Nizhni Novgorod, 603950 Nizhny Novgorod, Russia
        \and
        Maurizio De Pitt\`a \\
            Department of Neurobiology, The University of Chicago, Chicago, IL 60637, USA\\
            EPI BEAGLE, INRIA Rh\^{o}ne-Alpes, 60097 Villeurbanne, France       
        }
\begin{document}
%% Title
\maketitle

\section{Introduction}

Although~Ca$^{2+}$ ions have been identified as a key component of astrocyte signaling in response to stimuli, and could mediate the astrocyte's modulatory effects on the surrounding neuropile \citep{VolterraMeldolesiRev2005,Araque_Neuron2014}, their functional role remains intensely debated. Arguably, one of the principal reasons for such debate is that astrocytic~Ca$^{2+}$ shows highly complex spatiotemporal behavior \citep{Bindocci_Science2017}. In response to agonists, like hormones or neurotransmitters, and spontaneously as well, the majority of astrocytes exhibit oscillations of intracellular~Ca$^{2+}$ \citep{Verkhratsky_NCR2012,Zorec_ASN2012}. These oscillations can be grouped into two major types: those that are dependent on periodic fluctuations of the cell membrane potential and are associated with periodic entry of~Ca$^{2+}$ through voltage-gated~Ca$^{2+}$ channels, and those that occur in the presence of a voltage clamp. Our focus here is on the latter type and, in particular, on the so-called process of~Ca$^{2+}$-induced~Ca$^{2+}$ release (CICR) from the astrocyte's endoplasmic reticulum stores, which depends on cytosolic concentration of the second messenger inositol 1,4,5-trisphosphate \citep{Verkhratsky_MCE2012}.

Two main types of~IP$_3$-mediated CICR are observed in astrocytes \citep{Volterra_NRN2014,Rusakov_NRN2015}: (i)~transient~Ca$^{2+}$ oscillations that are confined to their (primary) processes, and (ii)~Ca$^{2+}$ elevations propagating along these processes as regenerative~Ca$^{2+}$ waves, often reaching the cell soma and triggering whole-cell~Ca$^{2+}$ signaling \citep{Pasti1997,SulHaydon2004}. Remarkably this latter kind of response can even propagate to neighboring astrocytes, through gap junction channels, and give rise to intercellular~Ca$^{2+}$ waves \citep{Kuga_etal_JN2011}. Although it is likely that different mechanisms could be in place depending on regional, developmental and physiological conditions, all these mechanisms likely depend, to some extent, on two key factors: the precise, molecular machinery underpinning~IP$_3$ signaling \citep{ScemesRev2001}, and the underlying spatiotemporal dynamics of synaptic stimulation of the astrocyte network \citep{VolterraMeldolesiRev2005}.

Both production and degradation of~IP$_3$ depend on enzymes that are regulated by cytosolic~Ca$^{2+}$ \citep{Berridge_etal_NatRev2003}. These include~Ca$^{2+}$-dependent phsopholipase~C$\updelta$- (PLC$\updelta$-)mediated~IP$_3$ synthesis and~Ca$^{2+}$-dependent~IP$_3$ degradation by~IP$_3$ 3-kinase and by inositol polyphosphate 5-phosphatase \citep{ZhangMuallem1993,SimsAllbritton1998,RebecchiPentyala2000}. Production by PLC$\updelta$ occurs in a regenerative fashion providing necessary~IP$_3$ amounts to promote CICR against enzymatic degradation of~IP$_3$ or rapid~IP$_3$ dilution by intracellular diffusion \citep{HoferGiaume2002,Ullah_CellCa2006}. On the other hand, while the activity of the 3-kinase is stimulated by~Ca$^{2+}$, the 5-phsophatase is inhibited instead \citep{CommuniErneux2001}, so that different mechanisms of degradation exist depending on~Ca$^{2+}$ concentration. In turn, these different routes for~IP$_3$ degradation have the potential to interact with PLC-mediated production in a diverse fashion, possibly ensuing in variegated~IP$_3$ and~Ca$^{2+}$ dynamics \citep{HouartGoldbeter1999,Goldberg_etal_PCB2010,MatrosovKazantsev_Chaos2011}.

Synaptic activation also constitutes a further mechanism to trigger or modulate~IP$_3$ signaling of the astrocyte, insofar as astrocytic receptors, targeted by synaptically-released neurotransmitters, are often part of the family of $G_q$-coupled receptors which, upon activation, promote~IP$_3$ production by PLC$\upbeta$ \citep{ZurNiedenDeitmer_2006}. Clustering of these receptors at points of contact of synapses with astrocytic processes \citep{DiCastro_Volterra_NatNeurosci2011,Panatier_etal_Cell2011,Arizono_STKE2012} provides spatially confined sites of~IP$_3$ production, whose differential activation could result in rich spatiotemporal~IP$_3$ and~Ca$^{2+}$ dynamics \citep{Volterra_NRN2014}. Nevertheless, it remains a matter of investigation, how the stochastic arrival of synaptic inputs within the anatomical domain of an astrocyte could be concerted to generate temporally precise~Ca$^{2+}$ signals.

Intense modeling efforts have been devoted in recent years to understand the richness of possible biophysical mechanisms underlying~IP$_3$-triggered CICR-mediated spatiotemporal dynamics in astrocytes (see \citet{Dupont_WileyRev2014} for a recent review). One of such mechanisms - intracellular~IP$_3$ diffusion - coupled with PLC-mediated~IP$_3$ production, have been suggested to crucially account for the whole spectrum of~Ca$^{2+}$ signaling, ranging from regular~Ca$^{2+}$ oscillations, to asynchronous, sparse and chaotic spatiotemporal~Ca$^{2+}$ dynamics \citep{HouartGoldbeter1999,Ullah_CellCa2006,KangOthmer_Chaos2009,MatrosovKazantsev_Chaos2011}. From a modeling perspective, the functional role of these~IP$_3$ pathways are studied within the more general framework of bifurcation analysis of the dynamics of nonlinear systems, since~IP$_3$ and~Ca$^{2+}$ signaling represent a famed example of such systems being modeled by nonlinear differential equations \citep{Fall_Book2002}. In this perspective, here we use tools of bifurcation theory to characterize dynamics of astrocytic~IP$_3$ and~Ca$^{2+}$ for different~IP$_3$ regimes from a mathematical point of view. We do so following a bottom-up approach, starting from a compact, well-stirred astrocyte model to first identify characteristic~IP$_3$ pathways whereby~Ca$^{2+}$ (and~IP$_3$) dynamics ``bifurcate'', namely change from stable (constant) concentration levels, to oscillatory dynamics. Then we extend our analysis to the elemental case of two astrocytes, coupled by~IP$_3$ diffusion mediated by gap junction channels, putting emphasis on the mechanisms of emergence of chaotic oscillations. Finally, we complete our analysis discussing spatiotemporal~Ca$^{2+}$ dynamics in a spatially-extended astrocyte model, gaining insights on the possible physical mechanisms whereby random Ca$^{2+}$~generation could be orchestrated into robust, spatially-confined intracellular~Ca$^{2+}$ oscillations.

\section{Birth and death of calcium oscillations in a compact astrocyte model}

\subsection{Compact astrocyte modeling}
Compact astrocyte models are useful to study cell-averaged signals such as whole-cell or somatic~Ca$^{2+}$ signals recorded in the majority of experiments \citep{SchusterReview2002,FalckeRev2004}. These models assume the astrocyte to be ``well-stirred,'' so that the concentration of each species is homogeneous throughout. Thus, for example, we write $C$ for the concentration of free~Ca$^{2+}$ ions in the cytoplasm and note that it is $C = C(t)$, that is $C$ has no spatial dependence. In the following, in particular, we consider emergence (or death) of~Ca$^{2+}$ oscillations in a popular compact astrocyte model originally developed by \citet{DePitta_JOBP2009}. The model describes astrocytic~Ca$^{2+}$ signaling by~Ca$^{2+}$-induced~Ca$^{2+}$ release (CICR) from the endoplasmic reticulum (ER) stores to the cytoplasm in combination with intracellular IP$_3$ dynamics. The model consists of three ordinary differential equations: one for intracellular (cytosolic)~IP$_3$ ($I$), one for cytosolic~Ca$^{2+}$ ($C$), and a further one for~Ca$^{2+}$-mediated deinactivation ($h$) of~IP$_3$ receptor/Ca$^{2+}$~channels, whereby these channels turn available to trigger CICR again after its occurrence (see \textcolor{red}{Chapter~5}). In particular, cytosolic~Ca$^{2+}$ and the~IP$_3$R deinactivation are described by a set of Hodgkin-Huxley-like equations, according to the description originally introduced by \citet{LiRinzel1994}, and evolve according to \citep{HoferGiaume2002}
\begin{align}
	\der{C}{t}     &= J_r(C,h,I) + J_l(C) - J_p(C) + J_{m}(C,I) \label{eq:C}\\
	\der{h}{t}     &= \Omega_h(C,I)\rpar{h_\infty(C,I)-h} \label{eq:h}
\end{align}
where $J_r,\,J_l,\,J_p,\,J_m$ respectively denote the~IP$_3$R-mediated~Ca$^{2+}$-induced~Ca$^{2+}$-release from the ER ($J_c$), the~Ca$^{2+}$ leak from the ER ($J_l$), the~Ca$^{2+}$ uptake from the cytosol back to the~ER by serca-ER~Ca$^{2+}$/ATPase pumps ($J_p$) \citep{DePitta_JOBP2009} and the~Ca$^{2+}$ flux across the plasmalemma membrane by other Ca$^{2+}$/ATPase (PMCA) pumps and leak mechanisms ($J_m$) \citep{Ullah_CellCa2006}. These terms, together with the~IP$_3$R deinactivation rate ($\Omega_h$) and the steady-state deinactivation probability ($h_\infty$), are given by \citep{DupontGoldbeter1993,LiRinzel1994}
\begin{flalign}
	J_r(C,h,I)      &= \Omega_C \, m_\infty^3 h^3 \, (C_T-(1+\varrho_A)C)\\
	J_l(C)          &= \Omega_L \, (C_T-(1+\varrho_A)C)\\
	J_p(C)          &= O_P \Hill{2}{C}{K_P}\\
	J_m{C,I}        &= O_l + O_s\Hill{2}{I}{K_s} - \Omega_s C\\
	\Omega_h(C,I)   &= \frac{\Omega_{2}(I+d_1)+O_2(I+d_3)C}{I+d_3}\\
	m_\infty(C,I)   &= \Hill{1}{C}{d_{5}}\Hill{1}{I}{d_{1}} \label{eq:m-inf}\\
	h_{\infty}(C,I) &= d_{2}\frac{I + d_{1}}{d_2(I+d_1)+(I+d_3)C}
\end{flalign}
where $\Hill{n}{x}{K}$ denotes the sigmoid (Hill) function $x^n/(x^n + K^n)$. In the absence of external stimulation, cytosolic~IP$_3$ concentration is regulated by the complex~Ca$^{2+}$-modulated interplay of phospholipase~C$\updelta$-mediated endogenous production and degradation by~IP$_3$ 3-kinase (3K) and inositol polyphosphatase 5-phosphatase (5P) \citep{ZhangMuallem1993,SimsAllbritton1998,RebecchiPentyala2000}, and evolves according to the mass balance equation \citep{DePitta_JOBP2009}
\begin{align}
	\der{I}{t}     &= J_\delta(C,I) - J_{3K}(C,I) - J_{5P}(I) \label{eq:I}
\end{align}
where
\begin{flalign}
	J_\delta(C,I)     &= O_\delta \, \frac{\kappa_\delta}{\kappa_\delta + I}\Hill{2}{C}{K_\delta}\\
	J_{3K}(C)         &= O_{3K}\, \Hill{4}{C}{K_D}\Hill{1}{I}{K_3} \label{eq:J-3K} \\
	J_{5P}(I)         &= \Omega_{5P}\, I \label{eq:J-5P}
\end{flalign}
A detailed description of the parameters in the above equations along with their estimation is provided in~\textcolor{red}{Chapter~5}. \tabref{T:model-parameters} in the appendix further summarizes the parameter values hereafter adopted.

\subsection{Bifurcations of~Ca$^{2+}$ and~IP$_3$ equilibria}
We start our analysis of intracellular~Ca$^{2+}$ dynamics by
considering the bifurcations of $C,\,h,\,I$ equilibria, namely of
those situations where~Ca$^{2+}$,~IP$_3$ and~IP$_3$R
deinactivation are constant in the astrocyte. In doing so, we set
the rate of~IP$_3$ production by PLC$\updelta$ sufficiently small so
as not to have spontaneous oscillations for typical~Ca$^{2+}$
and~IP$_3$ resting concentrations, i.e. $<200$~$\upmu$M \citep{KangOthmer_Chaos2009}, and we map how many the equilibria are and how their stability changes for different regimes of~IP$_3$ degradation,
mimicked by different values of the two~IP$_3$ degradation rates
$O_{3K}$ and $\Omega_{5P}$ in equations~\ref{eq:I}--\ref{eq:J-5P}.

\begin{figure}[!tp]
\centering
\includegraphics[width=\textwidth]{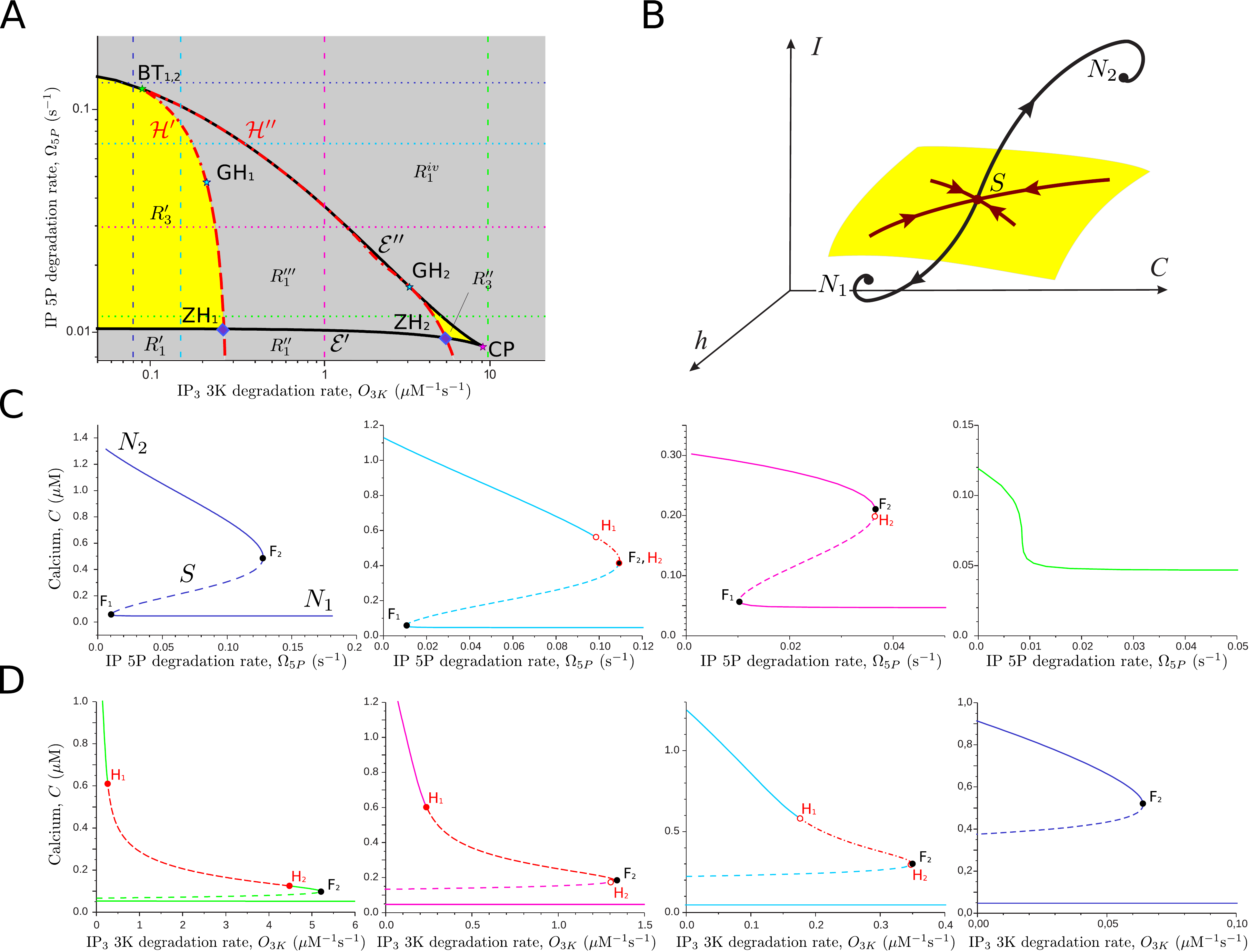}
\caption{Ca$^{2+}$~equilibria. \textbf{A}~Regions of existence for one (\textit{gray areas}, $R_1'-R_1^{iv}$) or two stable Ca$^{2+}$~concentrations in an astrocyte (\textit{yellow areas}, $R_3',\,R_3''$) for different rates of IP$_3$~degradation by IP$_3$~3-kinase ($O_{3K}$) and IP~5-phosphatase ($\Omega_{5P}$). The regions result from tiling of the parameter plane by fold (or saddle-node) ($\mathcal{E}',\,\mathcal{E}''$) and Andronov-Hopf bifurcation curves ($\mathcal{H}',\,\mathcal{H}''$; subcritical: \textit{dashed-dotted curves}; supercritical: \textit{dashed curves}) as detailed in the text. \textbf{B}~In the presence of bistability, a low ($N_1$) and a high stable intracellular Ca$^{2+}$~concentration ($N_2$) may be measured in the same cell, but the probability of observation of each can differ as it is set by the folding of the stable manifold (\textit{yellow surface}) of the saddle point ($S$) that separates the domains of attraction of the two stable concentrations. \textbf{C},~\textbf{D}~Bifurcation diagrams of Ca$^{2+}$~equilibria associated with vertical (\textit{dashed}) and horizontal (\textit{dotted}) sections of the bifurcation plane in panel~\textbf{A} (stable equilibria: \textit{solid lines}; saddle points: \textit{dashed lines}). It may be noted that, for some regimes of degradation, the upper branch of the diagram departing from the fold bifurcation $F_2$ that associates with the stable high Ca$^{2+}$~concentration ($N_2$), is split into two parts by two Andronov-Hopf bifurcation points ($H_{1,2}$). In between these points (\textit{red portions} in the diagrams), the high Ca$^{2+}$~state is no longer constant but rather oscillates. Model parameters as in \tabref{T:model-parameters}. \textbf{C}~Vertical sections for: $O_{3K}=0.08$~$\upmu$Ms$^{-1}$ (\textit{blue}); $O_{3K}=0.13$~$\upmu$Ms$^{-1}$ (\textit{cyan}); $O_{3K}=1.0$~$\upmu$Ms$^{-1}$ (\textit{magenta}); $O_{3K}=9.5$~$\upmu$Ms$^{-1}$ (\textit{green}). \textbf{D}~Horizontal sections for: $\Omega_{5P}=0.011$~s$^{-1}$ (\textit{green}); $\Omega_{5P}=0.03$~s$^{-1}$ (\textit{magenta}); $\Omega_{5P}=0.07$~s$^{-1}$ (\textit{cyan}); $\Omega_{5P}=0.135$~s$^{-1}$ (\textit{blue}).} \label{fig:bif-equilibria}
\end{figure}

\figref{fig:bif-equilibria}A summarizes the results of our
bifurcation analysis in the parameter plane $O_{3K}-\Omega_{5P}$.
In this plane, four bifurcation curves exist: two \textit{black curves} of fold (or saddle-node) bifurcation
points which originate at $O_{3K}\approx 8$~$\upmu$M by cusp
bifurcation (CP) and develop, for decreasing $O_{3K}$ values,
into two branches $\mathcal{E}',\,\mathcal{E}''$; and two
\textit{red curves} of Andronov-Hopf bifurcation points
$\mathcal{H}',\,\mathcal{H}''$, which are born through a pair of
consecutive Bogdanov-Takens bifurcations (BT$_{1,2}$) that are consistent with the 
existence of two zero eigenvalues for the system of
equations~\ref{eq:C}--\ref{eq:I}. Remarkably, the first Lyapunov
coefficient along the Andronov-Hopf curves changes from negative to positive
in coincidence of the two generalized Hopf points (GH$_1$, GH$_2$)
for increasing $O_{3K}$ values, marking a transition from
subcritical (\textit{dash-dotted red curves}) to supercritical
Andronov-Hopf bifurcations (\textit{dashed red curves}).

Fold and Andronov-Hopf curves also intersect in two
fold-Hopf points for low $\Omega_{5P}$ values -- $\mathcal{E}'$ with $\mathcal{H}'$ in ZH$_1$, and $\mathcal{E}''$ with $\mathcal{H}''$ in ZH$_2$ -- and ultimately partition the parameter plane into different regions which fall in two categories depending on the number of equilibria: (i)~one stable equilibrium, like in the four \textit{grey-shaded} regions $R_1'-R_1^{iv}$; and (ii)~two stable equilibria, as in the \textit{yellow-shaded} regions $R_3',\,R_3''$. Notably, in the case of monostability, the equilibrium is globally stable, that is~Ca$^{2+}$ and~IP$_3$ always converge to their equilibrium values from any initial state. In the presence of bistability instead, as schematically illustrated in~\figref{fig:bif-equilibria}B, the two equilibria $N_1,\,N_2$ are separated by a saddle point $S$ whose unstable manifold (\textit{yellow plane}) separates between the basins of attraction of the two equilibria.

Vertical and horizontal sections of the parameter plane
in~\figref{fig:bif-equilibria}A respectively provide codim-1
bifurcation diagrams in terms of the rate parameters $\Omega_{5P}$ (\figref{fig:bif-equilibria}C) and
$O_{3K}$ (\figref{fig:bif-equilibria}D), and illustrate different possible scenarios for
transitions from monostability to bistability and vice versa. In
particular, it may be noted that, for increasing rates of 3K-mediated IP$_3$
degradation (\figref{fig:bif-equilibria}C), the range of $\Omega_{5P}$ values for which bistability occurs, is mainly controlled by two fold bifurcations (F$_{1,2}$), unless BT$_1<\Omega_{5P}<$ZH$_1$ (\figref{fig:bif-equilibria}C, \textit{cyan diagram}) (or GH$_2<\Omega_{5P}<$ZH$_2$, not shown), in which case the high stable equilibrium ($N_2$) disappears via Andronov-Hopf bifurcation (H$_1$) as $\Omega_{5P}$ increases. Similar considerations also hold for the dual case where 5P-mediated~IP$_3$ is kept constant and $O_{3K}$ changes (\figref{fig:bif-equilibria}D). In this latter scenario however, as far as ZH$_1<\Omega_{5P}<$GH$_2$, the high Ca$^{2+}$ state can exists both for low and high $O_{3K}$ values (\figref{fig:bif-equilibria}D, \textit{green diagram}). In this case, low and high $O_{3K}$ values are separated by an unstable equilibrium comprised between two supercritical Andronov-Hopf points (H$_{1,2}$, \textit{dashed red line}) for which Ca$^{2+}$ oscillations emerge.

\subsection{Emergence of~Ca$^{2+}$ and~IP$_3$ oscillations and related limit cycle bifurcations}\label{sec:osc-1}
\begin{figure}[!tp]
\centering
\includegraphics[width=\textwidth]{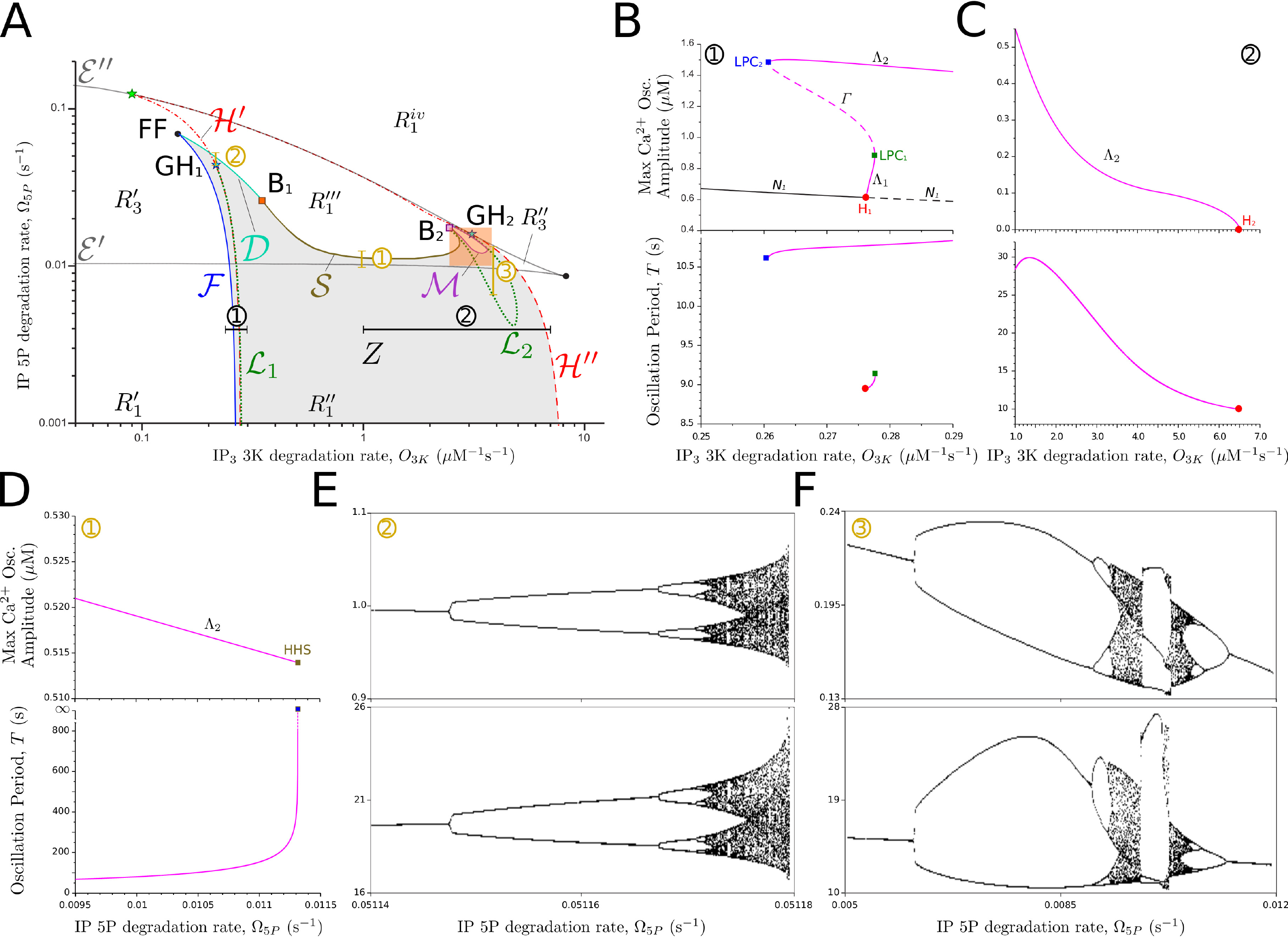}
\caption{Mechanisms of~Ca$^{2+}$ oscillations. \textbf{A}~The diagram in~\figref{fig:bif-equilibria}A is reproduced here with the addition of a \textit{gray-shaded area} $Z$, where rich oscillatory Ca$^{2+}$~dynamics exists for different regimes of IP$_3$~degradation. This region is bounded by a fold of cycles curve ($\mathcal{F}$) and a period doubling curve ($\mathcal{D}$) which originate by a fold-flip point (FF) for low rates of IP$_3$~degradation by 3-kinase ($O_{3K}$). For increasing $O_{3K}$ values instead, curve~$\mathcal{D}$ turns into a single-loop homoclinic-to-hyperbolic saddle curve ($\mathcal{S}$) at point B$_1$, which later becomes a multi-loop homoclinic-to-hyperbolic saddle curve ($\mathcal{M}$) via the bifurcation point B$_2$. Ca$^{2+}$~oscillations ultimately vanish for sufficiently high $O_{3K}$ values by a supercritical Andronov-Hopf bifurcation (\textit{dashed part} of the $\mathcal{H}''$ curve). \textbf{B},~\textbf{C}~Bifurcation diagrams for fixed rate of IP$_3$~degradation by~5P ($\Omega_{5P}=0.03$~s$^{-1}$) and increasing rates of 3K-mediated IP$_3$~degradation. Arbitrarily small Ca$^{2+}$~oscillations ($\Lambda_2$) can coexist with large-amplitude ones ($\Lambda_1$), independently of 5P-mediated degradation, for low $O_{3K}$ values (\textbf{B} and \textit{black segment} ``1'' in \textbf{A}) although their respective mechanisms of emergence/death are different (see text). \textbf{F}--\textbf{E} Bifurcation diagrams for fixed rates of IP$_3$~degradation by~3K and increasing rates of IP$_3$~degradation by~5P. It may be noted that, as $\Omega_{5P}$ approaches either the $\mathcal{D}$ or $\mathcal{M}$ curve, complex multirhythmic and/or chaotic oscillatory dynamics emerges (\textbf{D},~\textbf{E} and, respectively, \textit{yellow segments} ``2'' and ``3'' in~\textbf{A}). For sample dynamics within the \textit{orange-shaded rectangle}, see \figref{fig:osc-chaos-one}. \textbf{D}~$O_{3K}=1.0$~$\upmu$Ms$^{-1}$; \textbf{E}~$O_{3K}=0.21$~$\upmu$Ms$^{-1}$; \textbf{F}~$O_{3K}=4.0$~$\upmu$Ms$^{-1}$. Other model parameters as in \tabref{T:model-parameters}.}
\label{fig:bif-cycles}
\end{figure}

We now turn our analysis to the possible mechanisms underpinning
generation and death of~Ca$^{2+}$ and~IP$_3$ oscillations. From a
dynamical system perspective,~Ca$^{2+}$ and~IP$_3$ oscillations
are consistent with existence of limit cycle attractors in the
$O_{3K}-\Omega_{5P}$ plane, so that their birth (or death) is by
bifurcations of limit cycles. These bifurcations are shown in
\figref{fig:bif-cycles}A together with the bifurcation curves of
equilibria previously discussed (\figref{fig:bif-equilibria}A). In
this figure, five different bifurcation curves,
$\mathcal{F},\,\mathcal{D},\,\mathcal{S},\,\mathcal{M}$ and $\mathcal{H}''$, delimit a
\textit{gray-shaded} region $Z$ of the parameter plane
where~Ca$^{2+}$ and~IP$_3$ stable oscillations exist. In
particular, proceeding from left to right, one may note that curve
$\mathcal{F}$, which traces fold-of-cycles bifurcations, is born
with curve $\mathcal{D}$, which represents instead
period-doubling bifurcations, through a fold-flip bifurcation (FF). Curve $\mathcal{D}$ then turns
into a curve of single-loop homoclinic-to-hyperbolic saddle
bifurcations $\mathcal{S}$ at point~B$_1$. More precisely, this point corresponds to the single-loop separatrix of a saddle-focus bifurcation for which the saddle value of the associated homoclinic-to-hyperbolic saddle orbit is zero \citep{Kuznetsov1998}. An analogous point is also B$_2$, at higher $O_{3K}$ values, where the single-loop homoclinic orbit associated with $\mathcal{S}$ becomes multi-loop. This results in the existence of an infinite number of bifurcation lines in the $O_{3K}-\Omega_{5P}$ plane (not shown) which are bounded by curve $\mathcal{M}$ that marks the death of oscillations for increasing $\Omega_{5P}$ values. Curve $\mathcal{M}$ ultimately merges with the fold-of-cycle bifurcation curve $\mathcal{L}_2$ in proximity of the generalized Hopf point GH$_2$, whereby the \textit{dashed} supercritical Andronov-Hopf curve $\mathcal{H}''$ is born, and closes the oscillation region $Z$ for high $O_{3K}$ values.

\begin{figure}[!tp]
\centering
\includegraphics[width=\textwidth]{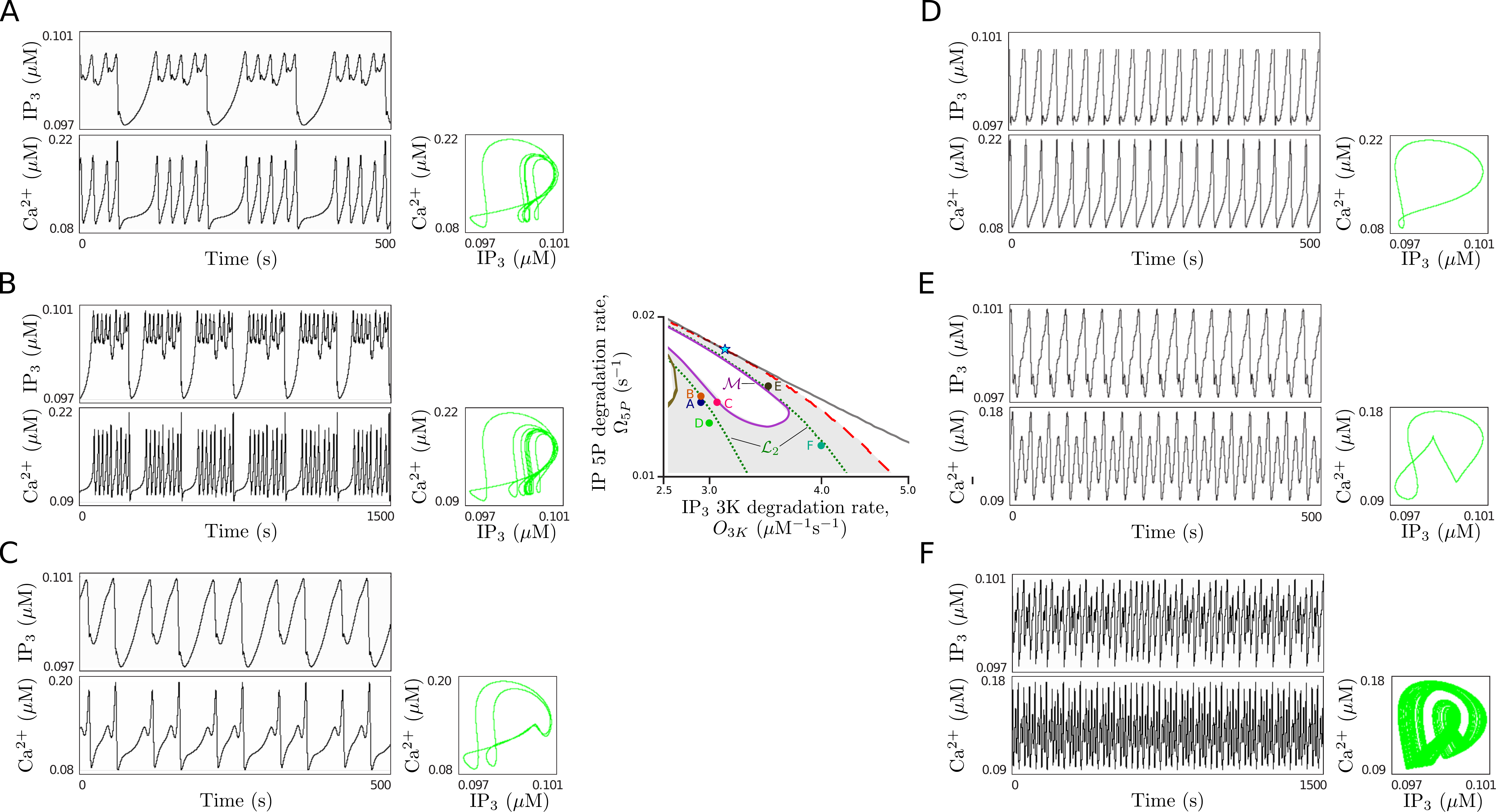}
\caption{Pathway to chaotic oscillations. \textbf{A}--\textbf{F}~Ca$^{2+}$~and IP$_3$~oscillations for different regimes of IP$_3$~degradation: each figure ensues from degradation rates marked by the corresponding labelled point in the plane $O_{3K}$ vs.~$\Omega_{5P}$ in the \textit{central panel} (zoom in on the \textit{orange-shaded rectangle} in \figref{fig:bif-cycles}A). Complex oscillations, including bursting, may be observed for IP$_3$~degradation regimes in the region delimited by the two curves $\mathcal{M}$ and $\mathcal{L}_2$ (\textbf{A}--\textbf{C},~\textbf{E}). These oscillations associate with Lissajous-like figures (in \textit{green}) that present a large, asymmetric ``8''-shaped curve, which maps the slow bursting component, and smaller concentric alike curves (occurring for higher Ca$^{2+}$~and IP$_3$~concentrations), that represent intraburst oscillations. On the contrary, for degradation rates outside this region, and sufficiently far from $\mathcal{L}_2$, simple oscillations exist which result in a single 8-shaped Lissajous curve (\textbf{D}). These oscillations can however rapidly turn complex (\textbf{E}) and/or chaotic (\textbf{F}) for degradation regimes approaching $\mathcal{L}_2$. \textbf{A}~$O_{3K}=2.939$~$\upmu$Ms$^{-1}$, $\Omega_{5P}=0.0135$~s$^{-1}$; \textbf{B}~$O_{3K}=2.939$~$\upmu$Ms$^{-1}$, $\Omega_{5P}=0.013513$~s$^{-1}$; \textbf{C}~$O_{3K}=3.26$~$\upmu$Ms$^{-1}$, $\Omega_{5P}=0.0135$~s$^{-1}$; \textbf{D}~$O_{3K}=3.0$~$\upmu$Ms$^{-1}$, $\Omega_{5P}=0.012$~s$^{-1}$; \textbf{E}~$O_{3K}=3.37$~$\upmu$Ms$^{-1}$, $\Omega_{5P}=0.014$~s$^{-1}$; \textbf{F}~$O_{3K}=4.0$~$\upmu$Ms$^{-1}$, $\Omega_{5P}=0.01055$~s$^{-1}$. Other model parameters as in \tabref{T:model-parameters}.}\label{fig:osc-chaos-one}
\end{figure}

Existence of such diverse bifurcations for different $O_{3K}$ and $\Omega_{5P}$ values, suggest different scenarios for birth and dynamics of~Ca$^{2+}$ and~IP$_3$ oscillations that depend on the regime of~IP$_3$ degradation. A first scenario of interest is the one found in correspondence of small $O_{3K}$ values, in an interval that includes the three curves $\mathcal{F},\,\mathcal{H}'$ and $\mathcal{L}_1$ (\textit{black segment}~``1'' in \figref{fig:bif-cycles}A). An inspection of the associated bifurcation diagram (\figref{fig:bif-cycles}B, \textit{top panel}) reveals that, starting from equilibrium~Ca$^{2+}$ and~IP$_3$ concentrations in region $R_1'$ (see \figref{fig:bif-equilibria}A),~Ca$^{2+}$ and~IP$_3$ oscillations of arbitrarily small amplitude ($\Lambda_1$) are born via a supercritical Andronov-Hopf bifurcation H$_1$ (which lies on $\mathcal{H}'$) for increasing $O_{3K}$ values. The limit cycle $\Lambda_1$ associated with these oscillations grows in amplitude with $O_{3K}$ in a small interval, till it disappears by the fold bifurcation LPC$_1$ (lying on $\mathcal{L}_1$). Nevertheless, oscillations may still be observed, yet of much larger amplitude, due to the presence of a large limit cycle $\Lambda_2$ which emerges via another fold bifurcation for $O_{3K}$ values in $R_1'$ (LPC$_2$, lying on $\mathcal{F}$). Thus, two scenarios of multistability exist: one for LPC$_2\le O_{3K} <$~H$_1$ and $O_{3K} >$~LPC$_1$, whereby constant (resting)~Ca$^{2+}$ and~IP$_3$ concentrations coexist with large oscillations of~Ca$^{2+}$ and~IP$_3$ oscillations; and another for H$_1 \le O_{3K} <$~LPC$_1$, where these large oscillations also coexist with smaller ones but there is no possible resting equilibrium for Ca$^{2+}$ and~IP$_3$. The ensuing astrocytic Ca$^{2+}$ dynamics, namely whether it is constant or oscillatory and, in this latter case, whether oscillations are small or large, ultimately depends on the location of the initial conditions (and thus on the astrocyte's history of activation) with respect to manifolds of the limit cycle $\Gamma$, which separate between the basins of attraction of the equilibrium ($N_1$) and the two limit cycles ($\Lambda_1,\,\Lambda_2$). It is also possible that both oscillations occur together resulting in bursting (not shown, but see \figref{fig:osc-chaos-one} for an example). In this case, as evidenced by the period of oscillations (\figref{fig:bif-cycles}B, \textit{bottom panel}), the small amplitude oscillations ($\Lambda_1$) are slower than larger ones, and thus set the inter-burst period, whereas intra-burst oscillations, which must be faster, are those of large amplitude associated with the limit cycle $\Lambda_2$.

The limit cycle $\Lambda_2$, on the other hand, exists across the whole $Z$ region, but its amplitude and period may considerably change depending on the IP$_3$~degradation regime under consideration. This is illustrated in \figref{fig:bif-cycles}C, where the maximal amplitude of Ca$^{2+}$~oscillations lying on $\Lambda_2$ are shown in correspondence of the same rate of 5P-mediated IP$_3$~degradation as in \figref{fig:bif-cycles}B, but faster rates of 3K-mediated IP$_3$~degradation (\textit{black segment}~``2'' in \figref{fig:bif-cycles}A). In this scenario, it may be noted that the cycle is maximally large for $O_{3K} \approx 1$~$\upmu$Ms$^{-1}$ but its amplitude is almost half of that observed in \figref{fig:bif-cycles}B, while the period is about 3-fold longer. In the transition from \figref{fig:bif-cycles}B to~\figref{fig:bif-cycles}C, as $O_{3K}$ increases beyond LPC$_1$, the nonlinear amplification by the Ca$^{2+}$-dependent Hill term in 3K-mediated IP$_3$~degradation (\eqref{eq:J-3K}) grows stronger reducing intracellular IP$_3$, thereby limiting availability of open IP$_3$Rs (\eqref{eq:m-inf}). This results in a weaker CICR from the ER stores which is reflected by smaller Ca$^{2+}$~oscillations. At the same time it also takes longer for IP$_3$~to reach the CICR threshold from baseline concentration values, which accounts for larger oscillation periods. Clearly, the effect is stronger with larger $O_{3K}$ values, and $\Lambda_2$ ultimately shrinks to arbitrarily small amplitude oscillations and disappears by supercritical Andronov-Hopf bifurcation in $H_1$ at $O_{3K}\approx 6.5$~$\upmu$Ms$^{-1}$.

Consider now the possible emergence of oscillations by any of the
three curves $\mathcal{D},\,\mathcal{S}$ and
$\mathcal{M}$. Among these three curves,
$\mathcal{S}$ is remarkable because allows rise or death of
regular (i.e. single-loop) oscillations ($\Lambda_2$,
\figref{fig:bif-cycles}D) with arbitrarily large period through
homoclinic-to-hyperbolic saddle bifurcation (HHS). Here, the saddle originates by a fold (or saddle-node)
bifurcation F$_1$ lying on $\mathcal{E}'$ and occurring at lower
$\Omega_{5P}$ (\textit{yellow segment}~``1'' in
\figref{fig:bif-cycles}A), so that the limit cycle coexist with the low equilibrium ($N_1$ in the \textit{third panel} of \figref{fig:bif-equilibria}C) for any value of $\Omega_{5P}$ comprised
between $\mathcal{E}'$ and $\mathcal{S}$.

Crossing of $\mathcal{D}$ and $\mathcal{M}$ may lead
instead to rich oscillatory dynamics. In the case of
$\mathcal{D}$ for example (\textit{yellow segment}~``2'' in
\figref{fig:bif-cycles}A and \figref{fig:bif-cycles}D), for increasing rates of degradation by
either 5-phosphatase or 3-kinase or both, regular~Ca$^{2+}$
and~IP$_3$ oscillations become irregular and eventually chaotic
through a typical period doubling cascade sequence \citep{Shilnikov_Book2001}.
A similar pathway to chaos may also be observed for $\Omega_{5P}$
approaching $\mathcal{M}$ (\textit{yellow segment}~``3'' in
\figref{fig:bif-cycles}A). In particular
as $\Omega_{5P}$ increases, first chaotic oscillations abruptly
vanish via intermittency (type 1) to later re-emerge through
period doubling cascade sequence, and finally become regular by
period halving cascade sequence. This complex dynamics ensues from the infinity of bifurcation curves that is found in proximity of $\mathcal{M}$ and which are omitted from the bifurcation portrait of \figref{fig:bif-cycles}A for obvious graphical reasons.

It suffices for the reminder of our analysis, to note that such complex dynamics can be observed in proximity of $\mathcal{M}$ and, in particular, for degradation regimes in the region bounded by this latter curve and the fold-of-cycles bifurcation curve $\mathcal{L}_2$, as approximately marked by the \textit{orange-shaded rectangle} in \figref{fig:bif-cycles}A. A zoom in on this region is considered in \figref{fig:osc-chaos-one} (\textit{central panel}) where sample Ca$^{2+}$~and IP$_3$~oscillations are shown for each labeled point therein. Considering for example the regimes marked by points A--C, it may be noted how small variations of the degradation rates are sufficient to dramatically alter shape and frequency of oscillations. In all cases, Ca$^{2+}$~and IP$_3$~oscillate in a bursting fashion with a constant $45^{\circ}-175^{\circ}$ phase shift, as reflected by the Lissajous-like curves in \textit{green}, but both shape and frequency of their intra- and inter-burst oscillations dramatically differ. Starting for example from \figref{fig:osc-chaos-one}A, a small increase of $\Omega_{5P}$, such as in the scenario of \figref{fig:osc-chaos-one}B, is sufficient to almost double duration and number of intraburst oscillations while slowing down overall occurrence of bursts (i.e. slower interburst  oscillations). The opposite instead occurs for a small increase of $O_{3K}$ whereby the frequency of bursts increases, but intraburst Ca$^{2+}$~oscillations almost tend to disappear in favor of more complex oscillations (\figref{fig:osc-chaos-one}C,E). While Ca$^{2+}$~bursting requires coexistence of at least two limit cycles as earlier noted (e.g.~\figref{fig:bif-cycles}B), and thus cannot be observed for IP$_3$~degradation regimes outside curve $\mathcal{L}_2$ (\figref{fig:osc-chaos-one}D), it also happens that, for regimes in proximity of this latter curve, the oscillations can become chaotic and highly irregular in shape and frequency (\figref{fig:osc-chaos-one}F).
 
To conclude, the possibility for existence of different~Ca$^{2+}$ and~IP$_3$ oscillatory dynamics depending on the regime of~IP$_3$ degradation is remarkable as it suggests different modes of encoding of stimuli by the astrocyte. In this fashion, periodic oscillations such as those in \figref{fig:osc-chaos-one}D,E could represent a mechanism of frequency encoding as far as their frequency, but not their shape nor their amplitude, changes with the stimulus. Conversely,~Ca$^{2+}$ bursting or chaotic oscillations (\figref{fig:osc-chaos-one}A--C,F), could perform more complex encoding, carrying stimulus information both in their frequency and amplitude, ultimately triggering different downstream effects \citep{DePitta_PRE2008,DePitta_JOBP2009}.

Noteworthy is that, in our description, chaotic oscillations are more likely to appear for large rates of~IP$_3$ degradation by IP~5-phosphatase. In particular, experimental evidence suggests that this enzyme could mainly localize in proximity of the plasma membrane, differently from~IP$_3$ 3-kinase which seems preferentially deeply anchored in the cytoplasm \citep{RebecchiPentyala2000,Irvine_AER2006}. Because on the other hand, the cytoplasm-to-ER ratio increases across an astrocyte, from soma to processes \citep{Pivneva_CellCalcium2008}, we could hypothesize that so does the relative expression of~IP$_3$-3K vs. IP~5P. In this fashion, different regimes of~IP$_3$ degradation could be present within different regions of an astrocyte: more 3K-driven ones in the processes at the cell's periphery, and others, where 5P contribution is stronger instead, in the soma or in primary processes branching from this latter. In turn, different regions of the astrocyte could differently encode stimuli, and their reciprocal disposition within the astrocyte anatomical domain, and with respect to the surrounding neuropile, could ultimately be correlated, in support of the possibility of subcellular organization of~Ca$^{2+}$ and~IP$_3$ signaling within the same cell \citep{Volterra_NRN2014}.

\section{IP$_3$ diffusion and regulation of~Ca$^{2+}$ and~IP$_3$ oscillations in connected astrocytic compartments}

\subsection{Modeling of astrocytic ensembles}
Either~Ca$^{2+}$ and~IP$_3$ signaling could extend to the whole
astrocyte or just be confined within a subcellular region of this
latter, our hitherto analysis has not taken into account the fact
that~IP$_3$ is highly diffusible in the cytoplasm. This could alter~IP$_3$ balance at the CICR site
(\eqref{eq:I}), either by subtracting or by adding~IP$_3$, with the
potential to affect~Ca$^{2+}$ signaling \citep{Sneyd_FASEB1995}. Remarkably
this scenario could hold either for subcellular propagation of~Ca$^{2+}$ waves mediated by
intracellular~IP$_3$ diffusion between neighboring subcellular
regions within the same cell, or for intercellular propagation of
whole-cell~Ca$^{2+}$ signals in networks of astrocytes connected by~IP$_3$-permeable gap junction channels \citep{ScemesRev2001}. Because~Ca$^{2+}$ waves are an important mechanism
whereby astrocytes could coordinate their behavior with that of
neighboring cells, and such waves often travel in an oscillatory manner
\citep{DiCastro_Volterra_NatNeurosci2011,Kuga_etal_JN2011} forming periodic waves, it is the period and the shape of these waves that are thought to control and coordinate a variety of cellular processes \citep{Verkhratsky_MCE2012}.

The mechanisms controlling period and shape of~Ca$^{2+}$ waves,
both intra- and intercellularly, are not well understood however
\citep{Bazargani_NN2016}. Thus, we devote this section to study how these wave
features could be regulated by~IP$_3$ signaling, and in particular
by~IP$_3$ diffusion. For this purpose, we consider a simple
description of oscillatory~Ca$^{2+}$ waves consisting of two
astrocytic compartments coupled by linear~IP$_3$ diffusion, where
by ``compartments'' we mean either two different astrocytes, or
two neighboring astrocytic regions. Accordingly, we
assume that~IP$_3$ diffusion is respectively
inter- or intra-cellular.

Each astrocytic compartment in our model is described by the same three variables $I,\,C,\,h$ used in the previous section. In particular, we keep equations~\ref{eq:C} and~\ref{eq:h} for $C$ and $h$ dynamics while we consider a somewhat simplified description of intracellular~IP$_3$ with respect to \eqref{eq:I}, so as to ease our analysis of the different contributions of~IP$_3$ production, degradation and diffusion in the emergence of~Ca$^{2+}$ waves \citep{Kazantsev2009,MatrosovKazantsev_Chaos2011}. In particular, we neglect~IP$_3$ competitive inhibition on PLC$\updelta$, as this mechanism is known to only marginally affect~IP$_3$ signaling for $I\gg \kappa_\delta$ \citep{DePitta_JOBP2009}, and assume no cooperativity for the binding reaction of~Ca$^{2+}$ with this enzyme. Moreover we linearly scale the dependence of PLC$\updelta$ activation by~Ca$^{2+}$ according to \citep{DeYoungKeizerPNAS1992,Kazantsev2009}
\begin{equation}
	J_\delta = O_\delta \rpar{\Hill{1}{C}{K_\delta} + (1-\alpha)\Hill{1}{K_\delta}{C}} \label{eq:J_delta}
\end{equation}
where $\alpha$ controls~Ca$^{2+}$-dependent PLC$\updelta$ activation such that for $0\le \alpha \le 1$ it is $O_\delta \Hill{1}{C}{K_\delta} \le J_\delta \le O_\delta$. We also neglect~Ca$^{2+}$ dependence of~IP$_3$ degradation and assume that~IP$_3$ levels in our description are low, at most close to~IP$_3$-3K binding affinity for~IP$_3$, i.e. $I\le K_{3K}$, so that $J_{3K} \approx O_{3K} I / K_{3K}$ (\eqref{eq:I}). In this fashion, we are able to lump~IP$_3$ degradation into a single term $J_{deg}$ that linearly depends on~IP$_3$, i.e.
\begin{equation}
	J_{deg} = J_{3K} + J_{5P} \approx \frac{O_{3K}}{K_{3K}}I + \Omega_{5P}I = \Omega_I\,I
\end{equation}
where we defined $\Omega_I=\Omega_{5P} + O_{3K}/K_{3K}$ as the rate of~IP$_3$ degradation. Finally, we consider Fick's first law of diffusion to describe~IP$_3$ diffusion, so that \citep{Crank_B1979}
\begin{equation}
	J_{diff} = - D \Delta I
\end{equation}
where $D$ is the~IP$_3$ diffusion rate and $\Delta I$ reflects the gradient of~IP$_3$ concentration across the boundary of the astrocytic compartment, moving from inside to outside this latter. In this fashion, coupling two astrocytic compartments together by linear~IP$_3$ diffusion results in the following system of six differential equations \citep{MatrosovKazantsev_Chaos2011}:
\begin{align}
	\der{C_i}{t}     &= J_r(C_i,h_i,I_i) + J_l(C_i) - J_p(C_i) + J_{m}(C_i,I_i)\label{eq:C-1d}\\
	\der{h_i}{t}     &= \Omega_h(C_i,I_i)\rpar{h_\infty(C_i,I_i)-h_i}          && i = 1,2 \label{eq:h-1d}\\
	\der{I_i}{t}     &= J_\delta(C_i) + J_{deg}(I_i) + J_{diff}(\Delta I_i) \label{eq:I-1d}
\end{align}
where $\Delta I_1 = I_1 - I_2 = -\Delta I_2$.

\subsection{Bifurcation analysis of two linearly-coupled astrocytic compartments}

\begin{figure}[!tp]
\centering
\includegraphics[width=\textwidth]{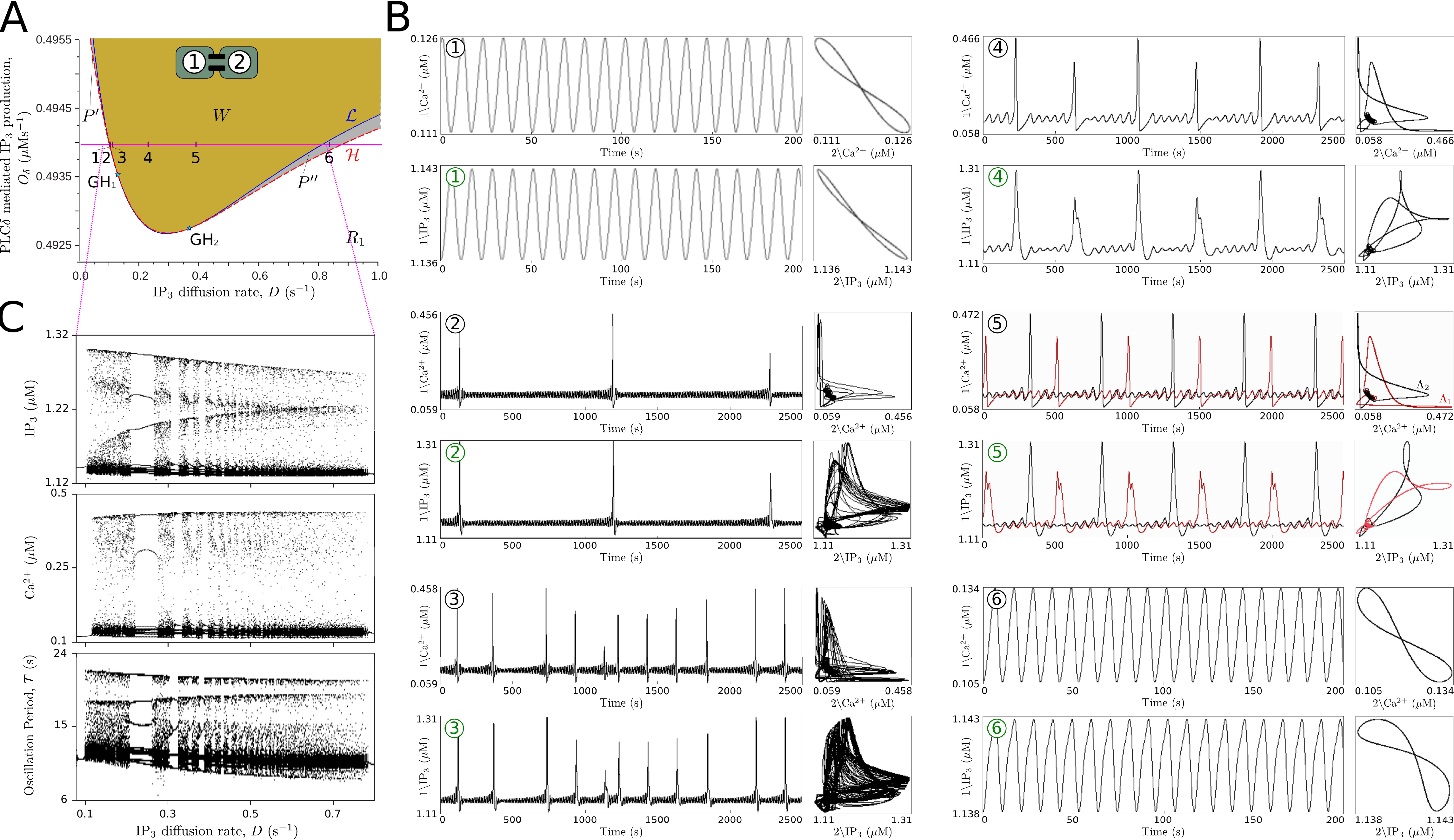}
\caption{Mechanisms of emergence of oscillations in two astrocytic compartments coupled by linear~IP$_3$ diffusion. \textbf{A}~For different rates of IP$_3$~production ($O_\delta$) and diffusion ($D$), Andronov-Hopf bifurcations ($\mathcal{H}$, \textit{red curve}) and fold of limit cycle bifurcations ($\mathcal{L}$, \textit{blue curve}) separate between regular (\textit{gray-shaded regions} $P',\,P''$) and chaotic Ca$^{2+}$ and IP$_3$~oscillations (\textit{yellow-shaded region}~$W$). Fixing the rate of IP$_3$~production by PLC$\updelta$ at $O_\delta=0.4939$~$\upmu$Ms$^{-1}$, different oscillations ensuing from different rates of IP$_3$~diffusion between the two compartments (\textit{points 1--6} on the \textit{magenta line} in~\textbf{A}) are shown in~\textbf{B}, along with the full bifurcation diagrams in function of IP$_3$~diffusion in~\textbf{C}. It may be noted from these diagrams, that both small and large-amplitude oscillations could coexist for~$D$ values within the $W$ region in~\textbf{A} (e.g.~panel \textbf{B}.\textbf{5}). Intermittent, quasi-periodic chaotic oscillations with decreasing frequency emerge for diffusion rates approaching the supercritical Andronov-Hopf bifurcations for $D<$~GH$_1$ or $D>$~GH$_2$ (\textit{dashed segments} of the \textit{red curve} $\mathcal{H}$ in~\textbf{A}), as shown in panels~\textbf{B}.\textbf{2} and \textbf{B}.\textbf{3} (corresponding to \textit{points~2} and~\textit{3} on the magenta line in~\textbf{A}). Regardless of the value of the IP$_3$~diffusion rate however, Lissajous curves ensuing from the time evolution of Ca$^{2+}$~and IP$_3$~dynamics in the two compartments (\textit{square panels} in~\textbf{B}) reveal that oscillations in the two compartments tend to be in phase opposition.~\textbf{B}.\textbf{1}:~$D=0.10562$~s$^{-1}$; \textbf{B}.\textbf{2}:~$D=0.106$~s$^{-1}$, \textbf{B}.\textbf{3}:~$D=0.12$~s$^{-1}$; \textbf{B}.\textbf{4}:~$D=0.25$~s$^{-1}$; \textbf{B}.\textbf{5}:~$D=0.3842$~s$^{-1}$; \textbf{B}.\textbf{6}:~$D=0.788$~s$^{-1}$. Other parameters as in \tabref{T:model-parameters}. In all simulations IP$_3$ degradation rate was fixed at $\Omega_I=0.1349$~s$^{-1}$. Because of symmetry, bifurcation diagrams in~\textbf{C} for quantities in either compartments are identical.}\label{fig:two-comp}
\end{figure}

\figref{fig:two-comp}A reveals existence of two bifurcations
curves for our system of two coupled identical astrocytic
compartments as a function of the~IP$_3$ diffusion rate ($D$) and
the~IP$_3$ production rate by PLC$\updelta$ ($O_\delta$): a
\textit{red curve} of Andronov-Hopf bifurcation points ($\mathcal{H}$), and
a \textit{blue curve} of fold bifurcation points of limit cycles
($\mathcal{L}$). These two curves intersect in two generalized Hopf bifurcation points
GH$_1$ and GH$_2$, in between of which Andronov-Hopf bifurcations are subcritical,
while being supercritical elsewhere. The parameter plane is
thereby subdivided into three regions: (i)~a \textit{white region}~$R_1$ where only constant (equilibrium)~Ca$^{2+}$ and~IP$_3$ concentrations are observed; (ii)~a \textit{grey-shaded} region ($P=P'\cup P''$) where regular~Ca$^{2+}$ and~IP$_3$ oscillations emerge; and
finally (iii)~a \textit{yellow region} $W$ where complex and/or chaotic oscillatory
dynamics occurs. Remarkably, oscillations either in $P$ or
in $W$ can only emerge for sufficiently large rates of
PLC$\updelta$-mediated~IP$_3$ production, independently of the rate
of~IP$_3$ diffusion, with a minimum value of $O_\delta$ for oscillations that depends on the rate of~IP$_3$
degradation (results not shown).

Based on the existence of the two curves $\mathcal{H}$ and
$\mathcal{L}$, two main mechanisms for birth (death) of
regular~Ca$^{2+}$ and~IP$_3$ oscillations may be expected:
one for $D<$~GH$_1$ and $D>$~GH$_2$, where oscillations of
arbitrarily small amplitude appear via supercritical Andronov-Hopf
bifurcation (Figure~\ref{fig:two-comp}B, \textit{panels}~1--6), and the other, for GH$_1<D<$~GH$_2$, whereby oscillations of arbitrarily small frequency emerge via subcritical Andronov-Hopf
bifurcation with amplitude set by the limit cycle previously born
through $\mathcal{L}$ (results not shown). Analysis of Lissajous-like curves associated
with the earlier scenario of oscillations (Figure~\ref{fig:two-comp}B, \textit{square panels}) show the
characteristic ``8''~shape of antiphase oscillations. This can be
explained noting that CICR in one compartment triggers a surge
of~IP$_3$ production by PLC$\updelta$ and an increase in diffusion
of~IP$_3$ from that compartment to the neighboring one, promoting
CICR therein. Yet, by the time this occurs,~Ca$^{2+}$-mediated
inactivation of~IP$_3$Rs in the first compartment has already
grown sufficiently to hinder CICR and make~Ca$^{2+}$ decrease, 
till~IP$_3$ diffusion from the other compartment
promotes~Ca$^{2+}$ increase again in a cyclic fashion, thereby
resulting in oscillations in the two compartments that are opposite
in phase (or almost so) \citep{Bindschadler_Chaos2001,Ullah_CellCa2006}.

The remarkable proximity between $\mathcal{H}$ and $\mathcal{L}$
curves, which respectively result in birth (or death) of local
vs.~global attractors in the parameter plane (i.e. equilibria vs.~limit cycles) \citep{Kuznetsov1998}, underpins further scenarios of emergence of oscillations that could not be observed in our previous bifurcation analysis of a single astrocytic compartment (Section~\ref{sec:osc-1}). In particular, for small~$D$ values increasing across the left branch of $\mathcal{L}$ (i.e.~for $D<$~GH$_1$),~Ca$^{2+}$
trajectories along the limit cycle emerging by $\mathcal{H}$ may
transiently enter the basin of attraction of the limit cycle
attractor emerging by $\mathcal{L}$ or vice versa, ensuing in
intermittent appearance of chaotic, pulse-like~Ca$^{2+}$ and~IP$_3$
oscillations (Figures~\ref{fig:two-comp}B.2,3) separated by
low-amplitude, quasi-periodic oscillations of variable duration.

Sampling of the bifurcation diagram in \figref{fig:two-comp}A for varying $D$ with constant $O_\delta$ (\textit{magenta segment}) indeed reveals existence of multiple chaotic bands for~$D$ values beyond $\mathcal{L}$, separated by intervals of almost regular oscillations whose amplitude decreases as~$D$ increases away from the $\mathcal{L}$ bifurcation boundary (\figref{fig:two-comp}C). These bands also reveals existence of two chaotic attractors, one for low and the other for high~Ca$^{2+}$/IP$_3$~values, as reflected by the two longitudinal bands in \figref{fig:two-comp}C. Notably, these two bands eventually merge for sufficiently large~$D$ values (and chaos disappears), ensuing in complex periodic oscillations which bear features of either or both attractors, that is either low- or large-amplitude oscillations (Figure~\ref{fig:two-comp}B.4) or both ($\Lambda_1$ and $\Lambda_2$ in Figure~\ref{fig:two-comp}B.5).

\section{Oscillatory Ca$^{2+}$~dynamics in the presence of stochastic IP$_3$~fluctuations}

\subsection{Spatially-extended astrocytic networks}
The case of two coupled astrocytic compartments considered in the previous section is arguably the simplest example of a ``spatially-extended'' astrocytic model, where space is defined by two points whose neighboring environment is approximated by a well-stirred compartment governed by equations~\ref{eq:C-1d}--\ref{eq:I-1d}. As previously mentioned, depending on the choice of model parameters, these two ``points'' (or compartments) can either represent two astrocytes connected by gap junctions, or two contiguous subcellular regions of the same cell. In either scenarios, we assume that the (Euclidean) distance between the two compartments is negligible with respect to their spatial extension, so that Ca$^{2+}$~and IP$_3$~dynamics in between the two compartments is merely a function of the concentrations of these species within the compartments. Remarkably, this can be extended to any number of compartments -- as far as properly well-stirred compartments are identified (see \textcolor{red}{Chapter~7}) -- so that the compartmental approach can be adopted to model both astrocytic networks \citep{Kazantsev2009,Goldberg_etal_PCB2010,Lallouette_FCN2014,Wallach_PCB2014} and individual astrocytes with spatially-extended, coarse-grained yet realistic geometry \citep{DePitta_PhDThesis,Wu_CellCalcium2014}.

What distinguishes whether our compartments represent individual cells or subcellular regions is a combination of any of the following factors: (i)~the choice of model parameters of each compartment, insofar as whole-cell (somatic) Ca$^{2+}$~signals are different in shape, duration and frequency with respect to Ca$^{2+}$~signals that are confined within astrocytic processes \citep{Bindocci_Science2017}; (ii)~how the compartments are connected, and (iii)~the nature of these connections, since different are the mechanisms regulating the IP$_3$~flow between compartments, depending on whether these latter represent cells or subcellular portions, for which nonlinear vs. linear coupling choices can be made accordingly \citep{Goldberg_etal_PCB2010,Lallouette_FCN2014}. On the other hand, there is emerging evidence suggesting that Ca$^{2+}$~signaling, both within a single astrocyte or in an astrocytic network, bears some degree of ``functional organization'' to the extent that, either different cellular regions \citep{DiCastro_Volterra_NatNeurosci2011,Bindocci_Science2017}, or portions of the network display unique spatiotemporal Ca$^{2+}$~dynamics \citep{Kuga_etal_JN2011,Sasaki_etal_CerebCortex2011}. This would mean that once we built a compartmental model of an astrocyte, or of an astrocytic network, we could \textit{a priori} predict that either individual compartments in our model, or specific ensembles of connected compartments, underpin generation of unique, stereotypical Ca$^{2+}$~patterns, namely that Ca$^{2+}$~signaling, ensuing from our simulations, presents some degree of \textit{functional} compartmentalization.

To understand what, in our hypothetical model, could be responsible for the emergence of such functional Ca$^{2+}$~compartmentalization, we could imagine to start considering identical compartments connected in some non-random fashion, and argue, somewhat trivially, that the very non-random topology of connections, along with the nature of those connections, are responsible, at least to some extent, to the emergence of unique spatiotemporal Ca$^{2+}$~patterns \citep{Lallouette_FCN2014}. There is nonetheless the complimentary possibility that spatiotemporal Ca$^{2+}$~organization could also emerge by intrinsic properties of the very biochemistry beyond IP$_3$-mediated CICR \citep{Volterra_NRN2014} -- a hypothesis which we are going to investigate for the remainder of this chapter. With this aim, we consider a $N$-by-$N$ lattice of identical astrocytic compartments, each compartment being linearly coupled to neighboring compartments, so that either~2~or 4~connections per compartment can be counted, depending on whether the compartment is at the borders of the lattice or far from it (\figref{fig:osc-stochastic}A). For sufficiently large $N$, the symmetry of this setup allows linking emergence of spatiotemporal Ca$^{2+}$~patterns that we are presumably going to observe in our simulations, exclusively with the nature of biochemical reactions underpinning astrocytic IP$_3$/Ca$^{2+}$~signaling. Furthermore, to rule out that pattern formation in our simulations could be caused by the nature of the initial perturbation or the choice of initial conditions used to ignite IP$_3$/Ca$^{2+}$~signaling, we set $\alpha=0$ in \eqref{eq:J_delta}, thus neglecting the nonlinear Ca$^{2+}$~dependency of PLC$\updelta$ for the sake of simplicity, while adding a further term for stochastic IP$_3$~production either by spontaneous PLC$\updelta$ activation \citep{LavrentovichHemkin_JTB2008} or by activation of PLC$\upbeta$ by stochastic synaptic inputs \citep{AguadoSoriano2002}. Accordingly, IP$_3$~production by PLC isoenzymes is described by \citep{Wu_CellCalcium2014}:
\begin{align}
	J_{prod}(t) &= O_\beta\, \mathcal{U}_\beta(0,1|t_k)\sum_k \updelta(t-t_k) + O_\delta\, \mathcal{U}_\delta(0,1|\vartheta_n)\sum_n \updelta(t-\vartheta_n) \label{eq:J_prod}
\end{align}
where $\mathcal{U}_x(0,1|t)$ (with $x=\beta,\,\delta$) denotes the generation of a random number at time $t$ withdrawn from the standard uniform distribution, and accounts for random modulations of the maximal rate of IP$_3$~production; $t_k$ and $\vartheta_n$ respectively stand instead for the instants of synaptically-evoked and spontaneous nucleation of IP$_3$/Ca$^{2+}$~spikes, which may be assumed to be Poisson distributed at first approximation, in agreement with experimental observations \citep{SoftkyKoch_JN1993,ShadlenNewsome_JN1998,Skupin_etal_BiophysJ2008}. Each astrocytic compartment in the lattice is thus described by \citep{Kazantsev2009}:
\begin{align}
	\der{C_{ij}}{t}     &= J_r(C_{ij},h_{ij},I_{ij}) + J_l(C_{ij}) - J_p(C_{ij}) + J_{m}(C_{ij},I_{ij})\label{eq:C-lattice}\\
	\der{h_{ij}}{t}     &= \Omega_h(C_{ij},I_{ij})\rpar{h_\infty(C_{ij},I_{ij})-h_{ij}}\\
	\der{I_{ij}}{t}     &= J_{prod}(t) + J_{deg}(I_{ij}) + J_{diff}(\Delta_{ij} I) \label{eq:I-lattice}
\end{align}
where $\Delta_{ij}$ is the discrete Laplace operator whereby
\begin{equation}
	\Delta_{ij} I=I_{i+1,j} + I_{i-1,j} + I_{i,j+1} + I_{i,j-1} - 4 I_{ij}
\end{equation}
with the indices $i,j = 1,\ldots,N$ denoting the discrete spatial coordinates in the $N\times N$ lattice.

\subsection{Ca$^{2+}$~wave propagation and functional organization}
\begin{figure}[!tp]
\centering
\includegraphics[width=\textwidth]{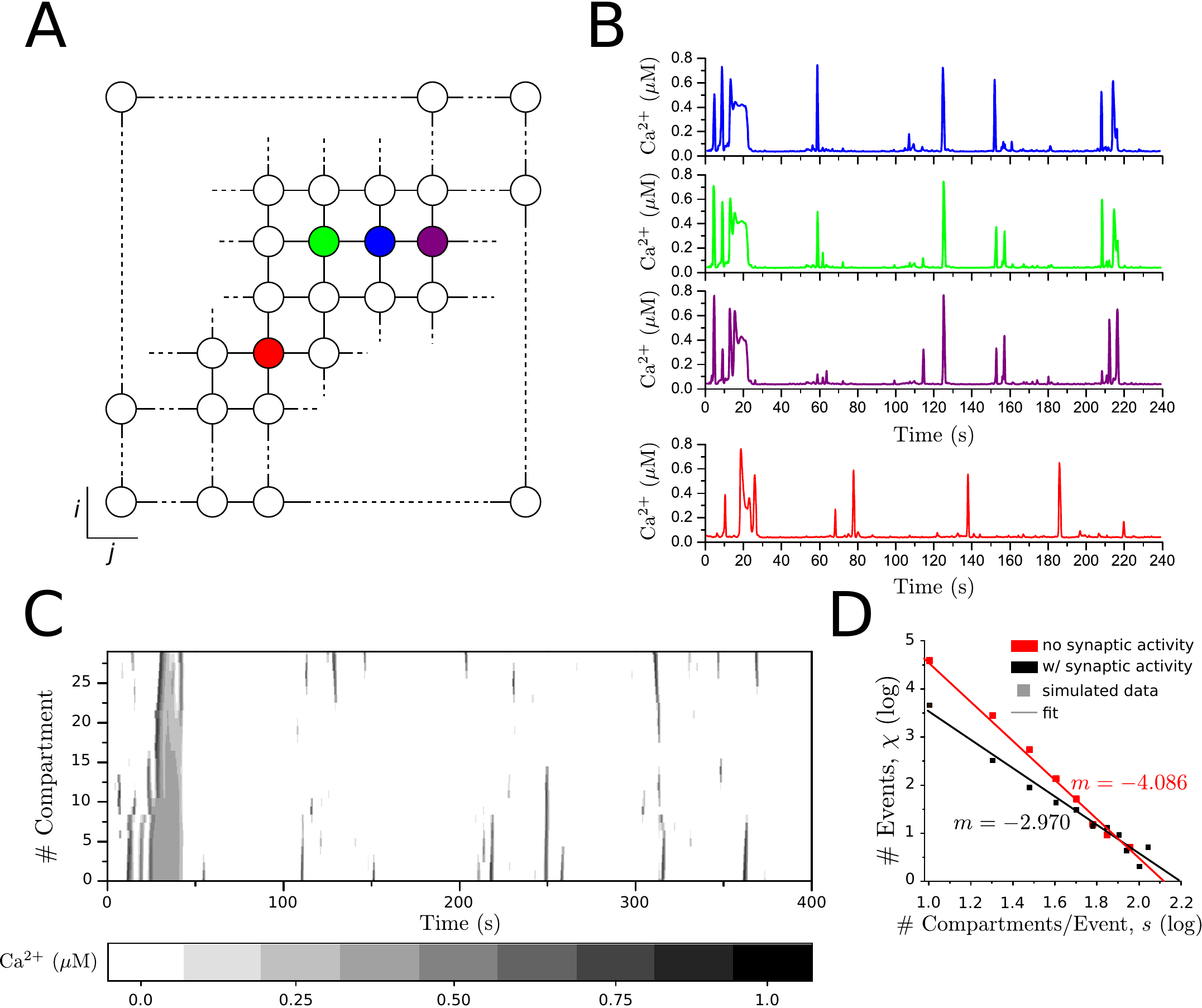}
\caption{Spatiotemporal~Ca$^{2+}$ dynamics emerging by stochastic IP$_3$/Ca$^{2+}$~nucleation. \textbf{A}~Illustration of the square lattice considered in the simulations for $30\times 30$ astrocytic compartments (\textit{circles}) connected with their nearest neighbors only. The \textit{colored compartments} associate with \textit{same colors} Ca$^{2+}$~traces in~\textbf{B} which were obtained for spontaneous IP$_3$~production mediated by PLC$\updelta$. \textbf{C}~Raster plot of Ca$^{2+}$~concentration dynamics for compartments in the first row of the lattice (\textbf{A}, $i=0$) reveals emergence of different spatiotemporal Ca$^{2+}$~patterns, ranging from frequent, spatially-confined Ca$^{2+}$~puffs encompassing $\sim$1--5 compartments, to sporadic larger Ca$^{2+}$~events comprising $>5$ compartments or even the whole ensemble of compartments (e.g. large \textit{gray} area for $t<50$). \textbf{D}~Log-log plot of the distribution of the number of Ca$^{2+}$~events over their size (in terms of number of active compartments per event) suggests a power law statistics hallmarking emergence of self-organized Ca$^{2+}$~patterns (\textit{solid line fits}). Remarkably, small Ca$^{2+}$~puffs observed for spontaneous IP$_3$/Ca$^{2+}$~nucleation by PLC$\updelta$ (\textit{red data points}) tend to disappear in the presence of synaptic activity (\textit{black data points}) in favor of the emergence of larger Ca$^{2+}$~events. Model parameters as in \tabref{T:model-parameters} except for \textbf{B},\textbf{C}:~$O_\beta = 0$, $O_\delta=12$~$\upmu$M, $\nu_\delta=10$~mHz; \textbf{D},~\textit{red points}:~$O_\beta = 0$, $O_\delta=12$~$\upmu$M, $\nu_\delta=10$~mHz; \textit{black points}:~$O_\beta = 5.5$~$\upmu$M, $\nu_\beta=0.1$~Hz; $O_\delta=0$. In all simulations: $\alpha = 0$, $\Omega_I=1$~s$^{-1}$ and $D=10$~s$^{-1}$.}
\label{fig:osc-stochastic}
\end{figure}

We first consider the scenario of spontaneous nucleation of Ca$^{2+}$~waves (i.e. $O_\beta = 0$) at some average rate $\nu_\delta$ in each compartment. Typical values for $\nu_\delta$ are in the range of~1--50~mHz \citep{Skupin_etal_BiophysJ2008,Bindocci_Science2017}, accordingly we chose an intermediate sample value of $\nu_\delta=10$~mHz. \figref{fig:osc-stochastic}B shows Ca$^{2+}$~traces sampled from different \textit{colored} compartments in the lattice in \figref{fig:osc-stochastic}A. It may be appreciated how Ca$^{2+}$~pulse-like fluctuations occur randomly and with variegated amplitudes in close analogy with experimental observations \citep{NettMcCarthy2002,Wu_CellCalcium2014}. These fluctuations generally present some degree of correlation in contiguous compartments as a result of linear IP$_3$~diffusion in between compartments (compare \textit{blue}, \textit{green} and \textit{purple} traces associated with same-color compartments in \figref{fig:osc-stochastic}A), but tend to be essentially uncorrelated for compartments far apart (see for example \textit{purple} and \textit{red traces}).

Looking at the raster plot of Ca$^{2+}$~activities in \figref{fig:osc-stochastic}C, built by stacking on the y-axis Ca$^{2+}$~traces of all $N=30$ compartments of the first row of the lattice ($i=0$ in \figref{fig:osc-stochastic}A), it may be seen how, despite spontaneous (random) IP$_3$/Ca$^{2+}$ pulse nucleation in each compartment, the ensemble of compartments under consideration displays somewhat non-random patterns of Ca$^{2+}$~activity. For example a large Ca$^{2+}$~wave originates in compartments 15--17 and propagates for $20 < t < 50$~s to neighboring compartments, ultimately engulfing the whole row and other portions of the lattice that are not shown. For $t>50$~s instead, a series of spatially-confined Ca$^{2+}$~puffs spanning few ($<5$) compartments alternate with larger Ca$^{2+}$~events encompassing several ($>5-10$) compartments.  

To characterize emergence of spatiotemporal Ca$^{2+}$~patterns in our lattice it is convenient to consider the distribution of the number of Ca$^{2+}$~events ($\chi$) over their size ($s$), quantified in terms of active compartments per event. Data points from our simulations are shown as \textit{red squares} in \figref{fig:osc-stochastic}D, where the choice of logarithmic scales on both axes reveals a linear regression fit for these data points characterized by a negative slope $m=-4.086$ (\textit{red line}) which is the hallmark of power law statistic, i.e. $\chi \propto s^{-|m|}$ \citep{Newman_CP2005}. Such statistics suggests that in our setup, spontaneous IP$_3$/Ca$^{2+}$~nucleation promotes formation of clusters of simultaneously active compartments of different sizes, with decreasing probability as the cluster size increases. These scale-free clusters of active compartments -- ``scale-free'' insofar as there is no specific size (i.e.~scale) for the observed clusters, but many sizes are possible -- confirm our original hypothesis that CICR molecular machinery itself promotes functional organization of Ca$^{2+}$~dynamics in astrocytic cells and networks. The formation of clusters of active astrocytic compartments may indeed be regarded as a fingerprint of emergence of highly functionally-connected cellular or subcellular regions, whose different local Ca$^{2+}$~dynamics are strongly correlated.

It is instructive in this context to also predict how such spontaneous functional organization of astrocytic compartments could change in the presence of synaptically-evoked IP$_3$~production. With this aim, we repeat our simulations assuming that random IP$_3$~production in \eqref{eq:J_prod} is by PLC$\upbeta$ rather than by PLC$\updelta$ (i.e.~$O_\delta=0$). Accordingly, we consider the scenario of synaptic stimuli ensuing from spontaneous neural firing in the range of $\sim$0.1--5~Hz \citep{SoftkyKoch_JN1993,Haider_Nature2013}, with an average synaptic release probability between $\sim$0.09 \citep{Schikorski1997} and $\sim$0.6 \citep{StevensWang_Neuron1995}, which sets effective rates $\nu_\beta$ of synaptically-evoked IP$_3$~production in the range of $\sim$0.01--3~Hz \citep{Destexhe_etal_Neurosci2001}. Then, choosing an intermediate value of $\nu_\beta=0.1$~Hz, we also consider a low maximal rate of IP$_3$~production to account for the observation that astrocytic Ca$^{2+}$~puffs by spontaneous synaptic activity are at most similar in size and intensity to spontaneous ones \citep{DiCastro_Volterra_NatNeurosci2011}. The resulting distribution of Ca$^{2+}$~events over their size (\textit{black points} in \figref{fig:osc-stochastic}D) is still consistent with power law statistics, but the power law exponent in this case is smaller ($m=-2.970$) with respect to the case of spontaneous PLC$\updelta$ mediated IP$_3$/Ca$^{2+}$~nucleation. 

To seek some insights on the possible implications of this result, recall that the exponent $m$~of the power law can be regarded as a measure of the weight of the tail of the distribution, that corresponds in our case to the emergence of larger functional Ca$^{2+}$~islands. Hence, the smaller $m$ is, the larger is the proportion of observed large Ca$^{2+}$~events. This is arguably what we expect by synaptic stimulation of astrocytes: namely that small random Ca$^{2+}$~puffs tend to disappear as synaptic activity increases, leaving the place to  Ca$^{2+}$~events that progressively encompass larger astrocytic areas in a concerted fashion \citep{Bindocci_Science2017} -- a phenomenon reminiscent of percolation in the context of reactive-diffusive media \citep{VanagEpstein_PRL2001}. Remarkably, it may be noted that independently of the scenario under consideration, the tails of the two distributions in \figref{fig:osc-stochastic}D tend to overlap, suggesting that large Ca$^{2+}$~events emerge with low probability fixed by the inherent properties of the molecular CICR machinery, regardless of the rate of random IP$_3$/Ca$^{2+}$~nucleation \citep{Wu_CellCalcium2014}.

\section*{Conclusions}
We have shown how, depending on differences in the underpinning regimes of IP$_3$~degradation, astrocytic Ca$^{2+}$~signaling could unfold into an incredible, dynamically rich repertoire of oscillations -- from simple periodic oscillations, to complex chaotic bursts and variegated chirps in transitions from different steady states. While this does not answer the question of the functional need for such rich dynamical repertoire by astrocytes, it nevertheless suggests that these cells are capable of implementing complex manipulations of stimuli by encoding these latter by multiple features of different Ca$^{2+}$~oscillations (e.g. Ca$^{2+}$~peaks, frequency, shape,$\ldots$) \citep{DePitta_PRE2008,DePitta_CognProc2009}. 

Two important predictions follow from the models and simulations hitherto discussed. First is the observation that in our simulations, it is often sufficient to slightly perturb IP$_3$~degradation rates by IP~5P or IP$_3$-3K to produce dramatic changes in the ensuing Ca$^{2+}$~dynamics. This could result on one hand in the emergence of periodic, self-sustained stable oscillations \citep{DePitta_JOBP2009}, reminiscent of biological clocks observed for example in other cell types such as hepatocytes \citep{Hofer_BJ1999}, cardiac myocytes \citep{Maltsev_BJ2011} and other biological systems \citep{Goldbeter_Book1997}. On the other hand, it suggests that, Ca$^{2+}$~dynamics could ensue from fine tuning of cellular properties of the astrocyte, possibly meant to deploy specific physiological needs \citep{Volterra_NRN2014}. Second is the consideration, that the molecular CICR machinery beyond the majority of observed astrocytic Ca$^{2+}$~signals, endows both the single astrocyte and networks of astrocytes by functional organization. Namely an astrocyte (or a network thereof) could dynamically deploy Ca$^{2+}$~signals of different spatial extension, each with its unique potential functional meaning. It is then plausible to think that heterogeneous expressions of enzymes for IP$_3$~production and degradation \citep{Irvine_AER2006}, which has not been taken into account in our compartmental approach, could further contribute to the emergence of such functional parcelization of astrocytic anatomical domains -- a prediction that is left for proof by future experimental and theoretical investigations.

%%%%%%%%%%%%%%%%%%%%%%%%%%%%%%%%%%%%%%%%%%%%%%%%%%%%%%%%%%%%%%%%%%%%%%%%%%%%%%%%%%%%%
%% Appendices
%%%%%%%%%%%%%%%%%%%%%%%%%%%%%%%%%%%%%%%%%%%%%%%%%%%%%%%%%%%%%%%%%%%%%%%%%%%%%%%%%%%%%
\begin{appendices}
\renewcommand\thetable{\thesection\arabic{table}}
\section{Numerical methods}
Numerical integration of models in equations~\ref{eq:C}--\ref{eq:I}, \ref{eq:C-1d}--\ref{eq:I-1d}, \ref{eq:C-lattice}--\ref{eq:I-lattice} was pursued by a fourth-order Runge-Kutta integration scheme with adaptive step size control and maximum step of~5~ms \citep{Press_Recipes1992}. To construct bifurcation diagrams we deployed custom code in Fortran and C++ for standard methods of numerical codim-1 and codim-2 continuations \citep{GuckenheimerHolmes1986,Kuznetsov1998}. For each value of the bifurcation parameter considered in a continuation, up to~150 peak values of~Ca$^{2+}$ and~IP$_3$ oscillations were stored, allowing to reliably detect multistability or chaotic oscillations. Regular vs.~chaotic regimes were distinguished by numerical evaluation of Lyapunov eigenvalue spectrum of trajectories on the identified attractor \citep{Shilnikov_Book2001}. Accordingly, the attractor was dubbed chaotic if the Lyapunov spectrum included at least one positive eigenvalue. The complexity of the attractor was instead estimated by numerical evaluation of its power spectrum \citep{Shilnikov_Book2001}. 

Stability of equilibria was pursued by numerical computation of the eigenvalue spectrum in the linearized model. Stability of limit cycles was instead assed by numerical estimation of Floquet multipliers of the corresponding fixed point in Poincar\'e sections according to classical methods of nonlinear dynamics theory \citep{GuckenheimerHolmes1986}.

In the study of~Ca$^{2+}$ propagation in astrocyte networks in \figref{fig:osc-stochastic}D, an astrocyte compartment was dubbed active if its intracellular~Ca$^{2+}$ increased beyond a threshold value of 0.4~$\upmu$M. Accordingly, an event was counted every time one or more connected compartments were simultaneously active in a time window of 0.1~s. Power-law fit of event number~($\chi$) vs. size~($s$) was pursued by linear fit of data points on log-log plots by Origin~8.0 software package (OriginLab Corp., Northampton MA).

\newpage
\section{Parameter values of the different models}
\begin{table}[!h]
\begin{tabularx}{\textwidth}{ l X c c }
\hline
Parameter&  Description & Value & Unit\\
\hline
\multicolumn{4}{c}{\textit{Compact single astrocyte model}}\\
\hline
$C_T$        & Total cell free Ca$^{2+}$ concentration          & 2               & $\upmu$M\\
$\varrho_A$  & Ratio between ER and cytosol volumes             & 0.185           & -- \\
$\Omega_C$   & Maximal rate of Ca$^{2+}$ release by IP$_3$Rs    & 6               & s$^{-1}$\\
$\Omega_L$   & Maximal rate of Ca$^{2+}$ leak from the ER       & 0.11            & s$^{-1}$\\
$O_{P}$      & Maximal rate of Ca$^{2+}$ uptake by SERCA pumps  & 2.2             & $\upmu$Ms$^{-1}$\\
$K_{P}$      & Ca$^{2+}$ affinity of SERCA pumps                & 0.1             & $\upmu$M\\
$d_1$        & IP$_3$ dissociation constant                     & 0.13            & $\upmu$M\\
$d_2$        & Ca$^{2+}$ inactivation dissociation constant     & 1.049           & $\upmu$M\\
$d_3$        & IP$_3$ dissociation constant                     & 0.9434          & $\upmu$M\\
$d_5$        & Ca$^{2+}$ activation dissociation constant       & 0.082           & $\upmu$M\\
$O_2$        & IP$_3$R  binding rate for Ca$^{2+}$ inhibition   & 0.1335          & $\upmu$M$^{-1}$s$^{-1}$\\
$\Omega_2$   & IP$_3$R  unbinding rate for Ca$^{2+}$ inhibition & 0.14            & s$^{-1}$\\
$O_{\delta}$ & Maximal rate of IP$_3$ production by PLC$\updelta$ & 0.15            & $\upmu$Ms$^{-1}$\\
$K_{\delta}$ & Ca$^{2+}$ affinity of PLC$\updelta$                & 0.5             & $\upmu$M\\
$k_{\delta}$ & Inhibition constant of PLC$\updelta$ by IP$_3$     & 1.0             & $\upmu$M\\
$O_{3K}$     & Maximal rate of IP$_3$ degradation by IP$_3$-3K  & free            & $\upmu$Ms$^{-1}$\\
$K_{D}$      & Ca$^{2+}$ affinity of IP$_3$-3K                  & 0.5             & $\upmu$M\\
$K_{3}$      & IP$_3$ affinity of IP$_3$-3K                     & 1.0             & $\upmu$M\\
$\Omega_{5P}$& Maximal rate of IP$_3$ degradation by IP~5P      & free            & s$^{-1}$\\
\hline
\multicolumn{4}{c}{\textit{Astrocytic compartment model}}\\
\hline
$O_l$        & Constant Ca$^{2+}$~influx by plasmalemma membrane      &0.025           & $\upmu$Ms$^{-1}$\\
$O_s$        & Maximal Ca$^{2+}$-dependent rate of Ca$^{2+}$~influx by PMCA&0.2             & $\upmu$Ms$^{-1}$\\
$K_s$        & Ca$^{2+}$~affinity of PMCA                             &1.0             & $\upmu$M\\
$\Omega_s$   & Maximal rate of Ca$^{2+}$~extrusion by PMCA            &0.5             & s$^{-1}$\\
$\alpha$     & Strength of Ca$^{2+}$~dependence of PLC$\updelta$        & 0.8            & --\\
$\Omega_{I}$ & Maximal rate of IP$_3$ degradation                     & 0.1349         & s$^{-1}$\\
$D$          & IP$_3$~diffusion rate                                  & free           & s$^{-1}$\\
\hline  
\end{tabularx}
\caption{All simulations discussed in this chapter use values of model parameters specified in the following Table. Simulation-specific values of model parameters as well as ``free'' parameters whose values are not reported in the table, are detailed in figure captions instead. For simplicity, we assumed $J_m=0$ in~\eqref{eq:C} as well as in the single-cell bifurcation analysis. Moreover, in the compartmental model and simulations in Figures~\ref{fig:two-comp} and~\ref{fig:osc-stochastic} we set the lower (resting) IP$_3$~equilibrium at $I_0 = 0.16$~$\upmu$M. This was achieved replacing the variable~$I$ by $I-I_0$ in equations~\ref{eq:I-1d} and~\ref{eq:I-lattice}.}
\label{T:model-parameters}
\end{table}
\end{appendices}

%%%%%%%%%%%%%%%%%%%%%%%%%%%%%%%%%%%%%%%%%%%%%%%%%%%%%%%%%%%%%%%%%%%%%%%%%%%%%
%% Figures and captions
%%%%%%%%%%%%%%%%%%%%%%%%%%%%%%%%%%%%%%%%%%%%%%%%%%%%%%%%%%%%%%%%%%%%%%%%%%%%%
%%%%%%%%%%%%%%%%%%%%%%%%%%%%%%%%%%%%%%%%%%%%%%%%%%%%%%%%%%%%%%%%%%%%%%%%%%%%%
%% Figure 1
%%%%%%%%%%%%%%%%%%%%%%%%%%%%%%%%%%%%%%%%%%%%%%%%%%%%%%%%%%%%%%%%%%%%%%%%%%%%%
\newpage
\clearpage

%%%%%%%%%%%%%%%%%%%%%%%%%%%%%%%%%%%%%%%%%%%%%%%%%%%%%%%%%%%%%%%%%%%%%%%%%%%%%
%% Figure 2
%%%%%%%%%%%%%%%%%%%%%%%%%%%%%%%%%%%%%%%%%%%%%%%%%%%%%%%%%%%%%%%%%%%%%%%%%%%%%
\newpage
\clearpage

%%%%%%%%%%%%%%%%%%%%%%%%%%%%%%%%%%%%%%%%%%%%%%%%%%%%%%%%%%%%%%%%%%%%%%%%%%%%%
%% Figure 3
%%%%%%%%%%%%%%%%%%%%%%%%%%%%%%%%%%%%%%%%%%%%%%%%%%%%%%%%%%%%%%%%%%%%%%%%%%%%%
\newpage
\clearpage

%%%%%%%%%%%%%%%%%%%%%%%%%%%%%%%%%%%%%%%%%%%%%%%%%%%%%%%%%%%%%%%%%%%%%%%%%%%%%
%% Figure 4
%%%%%%%%%%%%%%%%%%%%%%%%%%%%%%%%%%%%%%%%%%%%%%%%%%%%%%%%%%%%%%%%%%%%%%%%%%%%%
\newpage
\clearpage

%%%%%%%%%%%%%%%%%%%%%%%%%%%%%%%%%%%%%%%%%%%%%%%%%%%%%%%%%%%%%%%%%%%%%%%%%%%%%
%% Figure 5
%%%%%%%%%%%%%%%%%%%%%%%%%%%%%%%%%%%%%%%%%%%%%%%%%%%%%%%%%%%%%%%%%%%%%%%%%%%%%
\newpage
\clearpage

\newpage
\bibliography{./chapter6.bib}

\end{document}